\documentclass{article}
\usepackage{arxiv_ijcai17}

\usepackage{times}
\usepackage{amsmath}
\usepackage{amsthm}
\usepackage{arydshln}
\usepackage{caption}
\usepackage{cleveref}
\usepackage{extarrows}
\usepackage{mathtools}
\usepackage{subfig}
\usepackage{url}
\usepackage{tablefootnote}
\usepackage{threeparttable}
\usepackage{tikz}
\usetikzlibrary{fit,positioning,arrows,automata,calc}

\newcommand*\txtcircled[1]{\tikz[baseline=(char.base)]{
            \node[shape=circle,draw,inner sep=0.5pt] (char) {#1};}}

\title{Fast Change Point Detection on Dynamic Social Networks}
\author{Yu Wang$^*\quad$ Aniket Chakrabarti$^*\quad$ David Sivakoff$^\#\quad$ Srinivasan Parthasarathy$^*$\\
$*$ Department of Computer Science and Engineering, $\quad\#$ Department of Statistics\\
The Ohio State University, Columbus, Ohio, USA\\
Contact: wang.5205@osu.edu or srini@cse.ohio-state.edu}

\begin{document}

\maketitle

\begin{abstract}
  % abstract
%This paper studies the graph evolution mechanism with various growth models, which is represented as a hidden Markov model, and the detection of the changing of the models. Experiments show that the evolution mechanism of a graph can be captured by simply sampling a small portion of edges and tracking their (joint) distribution. It is also demonstrated that without knowing the ground truth, the (joint) distribution tracking technique can perform comparably to techniques requiring the ground truth knowledge in detecting the change of the evolution mechanism.

A number of real world problems in many domains (e.g. sociology, biology, political science and communication networks) can be modeled as dynamic networks with nodes representing entities of interest and edges representing interactions among the entities at different points in time. 
A common representation for such models is the snapshot model - where a network is defined at logical time-stamps. 
An important problem under this model is change point detection. 
In this work we devise an effective and efficient three-step-approach for detecting change points in dynamic networks under the snapshot model. 
%We demonstrate the utility of this simple algorithm versus the current state-of-the-art on both synthetic and real world networks. 
%several real world graphs drawn from political science, social network analysis and economics. 
Our algorithm achieves up to 9X speedup over the state-of-the-art while improving quality on both synthetic and real world networks.

\end{abstract}

\vspace{-0.2cm}
\section{Introduction}
% introduction
Dynamic network analysis is increasingly used in 
complex application domains ranging from social networks (Facebook
network evolution \cite{snap})
to biological networks (protein-protein interaction \cite{shih2012identifying}), from political science
(United Nations General Assembly voting network \cite{voeten2012data}) to
communication networks (Enron network \cite{klimt2004enron}). Such dynamic networks are often represented 
using the snapshot model. Under this model, every network
snapshot (represented by a graph)
is defined at a logical timestamp. Two questions of fundamental
importance are -- (i) how does a network evolve? (ii)
when does a network change significantly so as to arise
suspicion that something fundamentally different is
happening?

Various generative models~\cite{peixoto2015modeling,zhang2016random} have been proposed to address question (i) - to explain the evolution of a network. 
They study network evolution under certain generative models~\cite{erd6s1960evolution,karrer2011stochastic}. 
In reality, the generative model itself might change, as addressed in question (ii) above. 
Existing work~\cite{AkogluTK14,ranshous2015anomaly} use complex methods to detect such changes. 
One drawback of those delicate methods is that they are time-consuming, and hence often not scalable (in terms of both network size and number of snapshots). 
We seek to find an efficient and effective solution that can scale up both with network size and with number of snapshots. 

In this paper, we present a simple and efficient algorithm based on likelihood maximization to detect change points in dynamic networks under the snapshot model. 
We demonstrate the utility of our algorithm on both synthetic and real world networks drawn from political science (congressional voting, UN voting), 
and show that it outperforms two recent approaches (DeltaCon\cite{koutra2016deltacon}, and LetoChange\cite{peel2014detecting}) in terms of both quality and efficiency.
Our work has the following contributions: 
\begin{enumerate}
\item Our approach is general purpose -- it can accommodate various snapshot generative models (see~\Cref{table:edgeP}). 
\item We model network evolution as a first order Markov process and consequently our algorithm accounts for the temporal dependency while computing the dissimilarity between snapshots.
\item Our algorithm is efficient and has constant memory overhead that can be tuned by a user controlled parameter.
\end{enumerate} 
We extensively evaluate our approach on synthetic as well as real world networks and show that our approach is extremely efficient (both in performance and quality). 

%This paper is organized in the following way: \Cref{sect:RelatedWork} presents the background and related works; \Cref{sect:Problem} formulates the problem and \Cref{sect:Method} discusses the methodology; \Cref{sect:Experiment} evaluates the algorithm and \Cref{sect:Conclusion} concludes the paper.
\vspace{-0.2cm}

\iffalse
\begin{enumerate}
\item We develop an algorithm for detecting change points in a dynamic network. Our approach is model agnostic - not requiring 
knowing the type of the underlying generative model. A key aspect of our algorithm is simplicity, and hence is real time and has low constant memory overhead. 
%specific knowledge of the underlying generative process. A key aspect of our algorithm is simplicity, and hence is real time and has low constant memory overhead. 
\item Our approach accounts for the temporal dependency of network evolution, and propose an edge probability estimator which is unbiased even when there is temporal dependency. 
Along with detecting change points, our method can additionally estimate the extent of temporal dependency.
\item We extensively evaluate our approach on synthetic as well as real world datasets and show that our approach is extremely efficient (both in performance and quality).
\end{enumerate} 
\fi

\section{Related Work}
\label{sect:RelatedWork}
Ranshous et al. \cite{ranshous2015anomaly}, and Akoglu et al. \cite{AkogluTK14} recently survey network anomaly detection. 
Our change point detection problem is similar to Type 4, the ``Event and Change Detection'', of the former: 
given a network sequence, a dissimilarity scoring function, and a threshold, a change is flagged if the dissimilarity of two consecutive snapshots is above the threshold. 
We differ in that we assume there is a latent generation model governing the network dynamics, 
and we are trying to detect the change in the latent space, while they did not explicitly mention the latent generation model. 
Moreover, we consider the temporal dependency across the snapshots while no work in the surveys accounted for temporal dependency.

DeltaCon~\cite{koutra2016deltacon} uses a graph similarity-based~\cite{berlingerio2012netsimile} approach to detect change points in dynamic networks. 
It derives the features of a snapshot based on sociological theories. And the feature similarity of each consecutive snapshot pair is calculated. 
That work is model agnostic (has no assumption on the generation model of networks), and is the state-of-the-art in terms of efficiency. We compare our algorithm against this.

Moreno and Neville~\cite{moreno2013network}, Bridges et al. \cite{bridges2015multi} and Peel and Clauset~\cite{peel2014detecting} develop network hypothesis testing based approaches. 
The advantage is that one can get a p-value of the test, which quantifies the confidence of the conclusion. 
However, these approaches have two shortcomings: firstly, they need to assume a specific generation model of the networks (mKPGM, GBTER and GHRG respectively); 
secondly, they are extremely slow, mostly due to the bootstrapping for p-value calculation. La Fond et al.'s work \cite{la2014anomaly} can also generate a p-value. 
It is tested against DeltaCon without reporting running time and efficiency concern is also mentioned in the paper. 
These algorithms will not work in our setting where the detection is done real time under bounded memory constraints. 
We compare our model agnostic algorithm against \cite{peel2014detecting}.

The DAPPER heuristic~\cite{caceres2013temporal} proposes a similar edge probability estimator as ours. 
However, it does not consider the temporal dependency of snapshots. 
%, nor does it prove the unbiasness providing the temporal dependency. 
Moreover, it focuses on temporal scale determination while ours focuses on change point detection. 
Loglisci et al. \cite{loglisci2015relational} study change point detection on relational network using rule-based analysis. 
Our approach uses (hidden) parameter estimation instead of semantic rule to infer the structure. 
Li et al. \cite{li2016detecting} propose an online algorithm, and consider temporal dependency. 
The problem they study is different from ours in that they study information diffusion on network with fixed structure and use continuous time. 
A recent work by Zhang et al.~\cite{zhang2016random} also studies the dynamic network in a Markov chain setting. 
They focus on community detection while we focus on change point detection. 
\vspace{-0.3cm}

\section{Problem Formulation}
\label{sect:Problem}
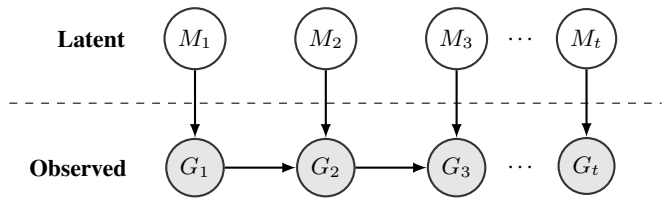
\begin{figure}[!t]
\vspace*{-0.0cm}
\begin{tikzpicture}
\footnotesize
\tikzstyle{main}=[circle, minimum size = 3.0mm, thick, draw =black!80, node distance = 9mm]
\tikzstyle{connect}=[-latex, thick]
\tikzstyle{box}=[rectangle, draw=black!100]
  \node[box,draw=white!100] (Latent) {\textbf{Latent}};
  \node[main] (M1) [right=0.4cm of Latent] {$M_1$};
  \node[main] (M2) [right=of M1] {$M_2$};
  \node[main] (M3) [right=of M2] {$M_3$};
  \node[main] (Mt) [right=of M3] {$M_t$};
  \node[main,fill=black!10] (G1) [below=of M1] {$G_1$};
  \node[main,fill=black!10] (G2) [right=of G1,below=of M2] {$G_2$};
  \node[main,fill=black!10] (G3) [right=of G2,below=of M3] {$G_3$};
  \node[main,fill=black!10] (Gt) [right=of G3,below=of Mt] {$G_t$};
  \node[box,draw=white!100] (Observed) [left=0.4cm of G1] {\textbf{Observed}};
  \path (M3) -- node[auto=false]{\ldots} (Mt);
  \path (G1) edge [connect] (G2)
        (G2) edge [connect] (G3)
        (G3) -- node[auto=false]{\ldots} (Gt);
  \path (M1) edge [connect] (G1);
  \path (M2) edge [connect] (G2);
  \path (M3) edge [connect] (G3);
  \path (Mt) edge [connect] (Gt);

  \draw [dashed, shorten >=-1cm, shorten <=-1cm]
    ($(Latent)!0.5!(Observed)$) coordinate (a) -- ($(Mt)!(a)!(Gt)$);

\end{tikzpicture}
\caption{\small Representation of the underlying generative process. Our inference is agnostic to $M_t$s.}
\label{fig:graphicalModel}
\vspace{-0.2cm}
\end{figure}

% model formulation
% problem formulation
This paper studies how to detect the times at which the fundamental evolution mechanism of a dynamic network changes. 
We assume that there is some \textit{unknown} underlying model that governs the generative process. 
Our change point detection algorithm is agnostic to this model. 
We assume that the observed network snapshots are samples that depend on some generative model and the previous snapshot. 
Networks have fluctuation across snapshots even when the generative model stays unchanged. 
Only when the generative model changes do we consider it a fundamental change. 
We represent the evolutionary process as a Markov Network (\Cref{fig:graphicalModel}).

\begin{table*}[!htb]
\small
\center
\caption{Edge probability between a dyad in each model}
\begin{tabular}{p{65pt}p{180pt}p{230pt}}
%\begin{tabular}{cll}
\hline\hline
Model & \hspace{20pt}Edge probability $ $ & Explanation\\
\hline
Erd\H os--R\'enyi (ER) & \hspace{20pt}$p(\langle n_{i},n_j\rangle\mid M)=p$ & $p$:  edge probability \\ \hdashline
Chung--Lu (CL) & \hspace{20pt}$p(\langle n_{i},n_j\rangle \mid M)=\beta w_iw_{j}/\sum_i w_i$ & $w_i$: weight of node $i$ (\cite{pfeiffer2012fast}); \newline $\beta$: edge density\\ \hdashline
Stochastic Block Model (SBM) & \hspace{20pt}$p(\langle n_{i},n_j\rangle\mid M)=p(c_i,c_{j})$ & $c_i$: community assignment of node $n_i$\newline  $p(r,s)$: probability of edges between communities $r$ and $s$\\ \hdashline
SBM-CL & \hspace{20pt}$p(\langle n_{i},n_j\rangle\mid M)\propto p(c_{i},c_{j})w_iw_{j}$ & notation as above\\ \hdashline
BTER & \hspace{20pt}$p(\langle n_{i},n_j\rangle\mid M)$\newline \textcolor{white}{space}$=p_{ER}I[c_i=c_j]+p_{CL}I[c_i\neq c_j]$ & Intra-community edge probability follows ER,\newline inter-community CL;~\cite{seshadhri2012community}\newline $I[\cdot]$ is the indicator function\\
\hline
\end{tabular}
\label{table:edgeP}
\vspace{-0.2cm}
\end{table*}

In \Cref{fig:graphicalModel}, 
 $M_t$ is the network generation model %random variable 
at time $t$. It is a triad $M_t=\langle\text{Type}_t,\Theta_t,\alpha_t\rangle$, where $\alpha_t$ is the continuity parameter at time $t$, 
Type$_t$ specifies the model, while $\Theta_t$ represents the model parameters (\Cref{table:edgeP} 
consists of some generative models we experiment on).
%we reiterate that our framework is generative-model agnostic, the table is necessarily inexhaustive). 
$G_t$ is the network (graph) observable at time $t$. 
We assume the number of vertices in $G_t$ is fixed to be $N$ for all snapshots 
(the union of all nodes is used when there is node addition/deletion, as in \cite{peel2014detecting}), 
so each $G_t$ has $2^{\binom{N}{2}}$ possible configurations, 
and $T$ is the total number of snapshots we observe. 
As per \Cref{fig:graphicalModel} the configuration of the network at time $t$, $G_t$, depends on the generation model at time $t$, 
$M_t$ (unobserved) and the network configuration at time $t-1$, $G_{t-1}$ (observed). 
Hence the networks in the observed sequence are samples from a conditional distribution (samples are not independent). 
The continuity rate parameter $\alpha_t$ controls the fraction of edges and non-edges that are retained from the previous snapshot, $G_{t-1}$. 
The network at time $t$ is assumed to be generated in the following way: for each dyad, 
keep the connection status from time $t-1$ with probability $1-\alpha_t$, and with probability $\alpha_t$, 
resample the connection according to the generation model at time $t$. 
Consequently, the smaller $\alpha_t$ is, the more overlap between two consecutive snapshots there is. 
Note that two consecutive network configurations may differ substantially if $\alpha_t>0$, even though the underlying generation model may be the same. 
Moreover, the changes of the generation model are assumed to be rare across the time span ($M_t \neq M_{t+1}$ is a rare event).

\begin{table}[!th]
\footnotesize
\caption{\small{Notation Table}}
\label{table:notationTable}
\begin{tabular}{p{1.1cm}|p{6.5cm}}
\hline\hline
Notation & Explanation\\
\hline
$N$ & network size, in terms of the number of nodes\\ \hdashline
  $T$ & number of snapshots\\ \hdashline
  $t$ & time stamp, $t\in\{1,\dots,T\}$\\ \hdashline
  $S$ & set of all change points\\ \hdashline
  $M_t$ & (unknown) generative model at time stamp $t$\\ \hdashline
  $G_t$ & snapshot at time stamp $t$\\ \hdashline
  $\alpha,\alpha_t$ & continuity rate (at time $t$)\\ \hdashline
  $s$ & window size, or number of snapshots in a window\\ \hdashline
  $W_t$ & a window of size $s$ ending at time $t$\\ \hdashline
  $\eta$ & step size of the sliding windows\\ \hdashline
  $w$ & number of windows: $\eta=1\rightarrow w=T-1;\;\eta=s\rightarrow w=\left\lceil T/s\right\rceil$\\ \hdashline
  $e_j$ & connection status of dyad $j$\\ \hdashline
  $\overline{M}$ & averaged number of edges in each snapshot: $\overline{M}=\sum^T_{t=1}\sum_{j=1}^{\binom{N}{2}}e_j^{(t)}/T$\\ \hdashline
  $p_j,p_j^{(t)}$ & connection probability of dyad $j$ (at time $t$)\\ \hdashline
  $k$ & number of dyads to be sampled and tracked\\ \hdashline
  $N_{01}^{(j)}$ & number of flips from 0 to 1 of dyad $j$ during the period of interest\\ \hdashline
  $N_{0*}^{(j)}$ & $N_{0*}^{(j)}\vcentcolon=N_{00}^{(j)}+N_{01}^{(j)}$, $N_{0*}^{(j)}+N_{1*}^{(j)}\equiv s-1$\\ \hdashline
  $N_{**}^{(j)}$ & $N_{00}^{(j)},N_{01}^{(j)},N_{10}^{(j)},N_{11}^{(j)}$\\ \hdashline
  $N_0^{(j)}$ & number of disconnected occurences of dyad $j$ in the period of interest\\
\hline
\end{tabular}
\end{table}

\noindent{\textbf{Problem Definition}} Our goal is to efficiently find a set $S\subset\{2,\dots,T\}$ such that $t\in S\iff M_t\neq M_{t-1}$, that is, to efficiently find all the time points at which the network generation model is different from the previous time point.
\vspace{-0.2cm}

\section{Methodology}
\label{sect:Method}
% methodology
Given the graphical formulation of the problem, 
exact inference is impossible since we do not know the underlying generative model, 
and our observations are stochastic.
However, even without prior knowledge of the generative model, 
we can still design an approximate inference technique based on MCMC sampling theory. 

%\subsection{Overview} 
The framework is straightforward, as mentioned in \Cref{sect:RelatedWork}: 
we first extract a ``feature vector'' from each snapshot, 
then quantify the dissimilarity between consecutive snapshots, 
and flag out a change point when the dissimilarity score is above a threshold. 
We use the joint edge probability as the ``feature vector'' (\Cref{subsect:Estimation}), 
exploit Kolmogorov-Smirnov statistic, Kullback-Leibler divergence and Euclidean distance for dissimilarity measure (\Cref{sect:DistanceM}), 
and use a permutation test like approach to determine the threshold (\Cref{subsect:Threshold}). 
\vspace{-0.2cm}

\subsection{Edge Probability Estimation} 
\label{subsect:Estimation}
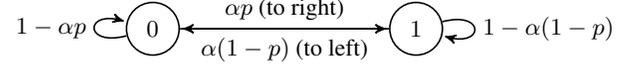
\begin{figure}
%\center
\hspace{0.3cm}
\footnotesize
\begin{tikzpicture}[->,>=stealth',shorten >=1pt,auto,node distance=3.5 cm, semithick, scale = 1.0, transform shape]

\node[state,minimum size=7mm] (0)              {$0$};
\node[state,minimum size=7mm] (1) [right of=0] {$1$};

\path (0) edge [above] node {$\alpha p$ (to right)} (1)
    (0) edge [loop left] node {$1-\alpha p$} (0)
    (1) edge [loop right] node {$1-\alpha(1-p)$} (1);
\path (1) edge [below] node {$\alpha(1-p)$ (to left)} (0);
\end{tikzpicture}
\caption{Two state Markov Chain}
\label{fig:MarkovChain}
\vspace*{-0.05in}
\end{figure}

In this subsection, we discuss how to (approximately) estimate the joint distribution of the dyads
\footnote{\footnotesize{we refer node pairs, which may or not be linked, as dyads}}. 
We track the presence or absence of a small fixed number of dyads throughout the entire observed sequence of network snapshots.
We break down the observation sequence into fixed-length windows, 
and for each window we infer the joint distribution of the dyads in our sample. 
We model each dyad to be a conditionally independent {\it two-state Markov chain \iffalse Edges are Markov chain, 
the generative model is encoded into the transition probability\fi} 
(\Cref{fig:MarkovChain}, we use $\alpha$ instead of $\alpha_t$ in this section for brevity) 
given the sequence of generative models $(M_t)_{t\ge 0}$ 
(this conditional independence assumption is satisfied for the choices of models in \Cref{table:edgeP}).
Note that even for generative processes that may result in greater dependence among dyads (such as the configuration model),
in many cases such dependence will be local, and if the number of dyads sampled is small, 
then these dyads will be spread out enough to be considered independent.
Moreover, the conditional independence assumption significantly improves computational efficiency (\cite{hunter2012computational}). 
The marginal probabilities of these dyads can then be estimated using the observed samples within each time window. 

We formalize the estimation procedure below. Given a network sequence $G_1,\dots,G_T$, 
we group the networks into sliding windows. 
We define $W_t$ to be a subsequence of $s$ consecutive observed networks ending at network $G_t$, so $W_t\equiv(G_{t-s+1},G_{t-s+2},\dots,G_t)$. 
We use equal sized sliding windows with a step size $\eta$, and we obtain a window sequence $W_s,W_{s+\eta},\ldots, W_{s+(i-1)\eta},\ldots,W_T$. 
Non-overlapping window setting uses $\eta=s$. 
In each window $i$, we can estimate the joint edge distribution (for the selected dyads) 
$P(e_1,e_2,\dots,e_k \mid M_{s+(i-1)\eta})$, where $e_j=1$ indicates an undirected edge between the $j$-th dyad, and $k<<\binom{N}{2}$ is the number of dyads tracked. 
For each of the models in \Cref{table:edgeP}, 
the joint distribution can be factorized into $P(e_1,e_2,\dots,e_k\mid M_{s+(i-1)\eta})=\Pi_{j}P(e_j\mid M_{s+(i-1)\eta})$. 
(conditional independence, see method description above) 

We can view a dyad across time as a two state Markov chain, 
and the chain length is the window size. We call a dyad across time a chain in the following text for brevity. 
Let $p_j\equiv P(e_j=1\mid M_{s+(i-1)\eta}),q_j\equiv1-p_j$, 
and suppose we are interested in $k$ chains. 

\smallskip
\noindent\textbf{1) Maximum Likelihood Estimator (MLE)}

The joint probability of the chains is (\Cref{fig:MarkovChain})
\begin{equation}
\footnotesize
\begin{multlined}
P(N_{**}^{(j=1,\dots,k)}\mid\alpha,\vec{p}) \\
=c_k\Pi_{j=1}^k(\alpha p_j)^{N_{01}^{(j)}}(\alpha q_j)^{N_{10}^{(j)}}(1-\alpha p_j)^{N_{00}^{(j)}}(1-\alpha q_j)^{N_{11}^{(j)}}
\end{multlined}
\end{equation}

%the likelihood function of $\alpha,\vec{p}$ given the observation $N_{**}^{(j)}$, 
where $N_{01}^{(j)}$ is the number of transitions from 0 to 1 for a chain 
(non-edge to edge for the dyad $j$ within the window), $c_k$ stands for the 
combinatorial coefficients independent of $\alpha,\vec{p}$. and $\sum_{p,q\in\{0,1\}}N_{pq}^{(j)}=s-1$ for 
all $j$s. And hence the log-likelihood (omitting the coefficient $c_k$) is:
\begin{equation}
\footnotesize
%\hspace*{-0.5in}
\begin{multlined}
L(\alpha,\vec{p}\mid N_{**}^{(j)})= \sum_{j=1}^k[N_{01}^{(j)}\ln(\alpha p_j)+N_{10}^{(j)}\ln(\alpha q_j) \\
+N_{00}^{(j)}\ln(1-\alpha p_j)+N_{11}^{(j)}\ln(1-\alpha q_j)]
%L(\alpha,\vec{p}\mid N_{**}^{(j)})=\sum_{j=1}^{\binom{N}{2}}[N_{01}^{(j)}\ln(\alpha p_j)+N_{10}^{(j)}\ln(\alpha q_j)+N_{00}^{(j)}\ln(1-\alpha p_j)+N_{11}^{(j)}\ln(1-\alpha q_j)]
\end{multlined}
\end{equation}

%One principled way is to estimate the parameters using MLE. 

\textbf{MLE for a single chain} First consider there is only one chain. 
Solving the zero-derivative \Cref{eq:Dp,eq:Dalpha}, leads to estimators of $\alpha,p$. 
And the estimators indeed lead to a negative definite Hessian, and therefore is the MLE. Hence we have

\begin{equation}\label{eq:EstOneChain}
\footnotesize
\hat\alpha_{\text{MLE}}=\frac{N_{01}N_{1*}+N_{10}N_{0*}}{N_{0*}N_{1*}}\qquad\hat p_{\text{MLE}}=\frac{N_{01}N_{1*}}{N_{01}N_{1*}+N_{10}N_{0*}}
\end{equation}

\textbf{MLE for multiple chains} The MLE for multiple chains essentially involves solving a high degree polynomial, which in general does not have a closed form solution.

\begin{equation}\label{eq:Dp}
\footnotesize
\frac{\partial L}{\partial p_j}=\frac{N_{01}^{(j)}}{p_j}-\frac{N_{10}^{(j)}}{q_j}-\frac{\alpha N_{00}^{(j)}}{1-\alpha p_j}+\frac{\alpha N_{11}^{(j)}}{1-\alpha q_j}=0
\end{equation}
\begin{equation}\label{eq:Dalpha}
\begin{aligned}
\frac{\partial L}{\partial\alpha}&=\sum\frac{N-N_{00}^{(j)}-N_{11}^{(j)}}{\alpha}-\frac{p_jN_{00}^{(j)}}{1-\alpha p_j}-\frac{q_jN_{11}^{(j)}}{1-\alpha q_j}=0\\
&\xLongrightarrow{\alpha\neq0}\sum\frac{N_{00}^{(j)}}{1-\alpha p_j}+\frac{N_{11}^{(j)}}{1-\alpha q_j}=kN
\end{aligned}
\end{equation}
where $1-\alpha$ is the continuity rate. If $\alpha=0$ then all snapshots are identical, which is uninteresting, so we have $\alpha\neq0$ in \Cref{eq:Dalpha}.

Combining \Cref{eq:Dp,eq:Dalpha}, one can get a high order polynomial of $p_j$, 
which in general does not have closed-form solutions by Abel-Ruffini theorem. 
We tried solving a special case where there are two chains by Wolfram Mathematica~\cite{mathematica}. 
The solutions (of two quartic functions) turn out to be very complicated and take over 40 pages. 
A common way to solve such maximization problems is to employ numerical methods such as gradient descent. 
The drawback of such an approach is that it can be computationally expensive with hundreds of dyads and windows. 
Therefore, we settle for an approximation of the MLE that empirically approximates numerical values well\footnote{\footnotesize{At significant level $0.05$, two sample $t$-test shows the approximated values equal to the numerical values.}}. 
Intuitively, the estimator for $\alpha$ should depend on all the chains, 
but chains that spend more time in both states $0$ and $1$ provide more information about $\alpha$ than chains that spend most time in one state 
(the latter may be due to small $\alpha$ or to a value of $p_j$ far from $1/2$). 
Since we can easily compute the MLE for $\alpha$ for a single chain, 
we estimate $\alpha$ with a weighted average of the MLEs from the individual chains, 
with chains that spend more time in both states being weighted more heavily.
We then estimate each $p_j$ by the MLE for the $j$th chain, since the chains are conditionally independent given $\alpha$. 
This results in the following estimators.

\vspace{-10pt}
\begin{equation}\label{eq:Approx}
\footnotesize
\vspace{-0cm}
\begin{multlined}
\hat\alpha_{\text{approx}}=\sum_jw_j\hat\alpha_{j\text{MLE}}=\sum_j w_j\frac{N_{01}^{(j)}N_{1*}^{(j)}+N_{10}^{(j)}N_{0*}^{(j)}}{N_{0*}^{(j)}N_{1*}^{(j)}} \\
\hat{p}_{j\text{approx}}=\hat{p}_{j\text{MLE}}=\frac{N_{01}^{(j)}N_{1*}^{(j)}}{N_{01}^{(j)}N_{1*}^{(j)}+N_{10}^{(j)}N_{0*}^{(j)}}\qquad\qquad\qquad
\end{multlined}
\end{equation}
\vspace{-0cm}
\footnotesize
\[\hspace{-0.4cm}\text{ where }w_j=\frac{[N_{0*}^{(j)}N_{1*}^{(j)}]^p}{\sum_j[N_{0*}^{(j)}N_{1*}^{(j)}]^p},\;\text{ and }\;N_{0*}^{(j)}=N_{00}^{(j)}+N_{01}^{(j)}\]
\normalsize

Empirically we find the exponent $p=\infty$ works best, 
which means $\hat\alpha_{\text{approx}}$ is a simple average of the $\alpha_{j\text{MLE}}$ corresponding to the chains with maximal value of $N_{0*}^{(j)}N_{1*}^{(j)}$.
The continuity rate describes the temporal dependency among networks, 
and can help us determine a proper window size. 

\textbf{Drawbacks of MLE} 
Though MLEs are consistent in general, there is no guarantee of unbiasness for these particular estimators with limited samples. 
%Though consistent in general, the above MLEs have no guarantee of unbiasness with limited samples. 
Moreover, they have three random quantities ($N_{01},N_{10},N_{1*},N_{0*}$ in (6) have 3 degrees of freedom for fixed $s$) and hence require more samples to estimate, making it prohibitive in practice.

\smallskip
\noindent\textbf{2) Simplified Estimator} 

To overcome the drawbacks of MLE, we propose a simple estimator for 
the edge probability which is consistent and unbiased, 
has only one random quantity and therefore requires fewer samples. 
The simple estimator essentially estimates the edge frequency in each window. 
If we know changes happen rarely, and the process stays in equilibrium in most of time, 
we can show the following estimator to be consistent and unbiased in equilibrium:
\begin{equation}
\footnotesize
\hat p_{j\text{eq}}\equiv\frac{N_{1*}^{(j)}+e_j^{[s+(i-1)\eta]}}{s}=\frac{\#\text{ of 1s in the chain}}{s}\equiv\frac{N_1^{(j)}}{s}
\end{equation}

which is the proportion of snapshots in which the dyad $j$ being an edge within the window. 

\noindent\textbf{Proposition 1} In equilibrium, $\hat{p}_{j\text{eq}}$ is consistent as chain length (window size) increases.
\vspace{-0.1cm}
\begin{proof}
$\pi^{(j)}_0\equiv P(\text{non-edge of chain }j\text{ in equilibrium}),\;s\equiv N^{(j)}_0+N^{(j)}_1(\text{fixed}),\;N^{(j)}_1\equiv\#\text{ of 1s in the chain}$

By ergodic theorem \cite{givens2012computational}, $\pi_1^{(j)}\xlongequal{\text{\normalsize almost surely}}\lim_{s\to\infty}N_1^{(j)}/s=\lim_{s\to\infty}\hat p_{j\text{eq}}$, 
where the first equation means almost sure convergence, and implies convergence in probability (estimator being consistent).
\end{proof}
\vspace{-0.1cm}
\noindent\textbf{Proposition 2} In equilibrium, $\hat{p}_{j\text{eq}}$ is unbiased.
\vspace{-0.3cm}
\begin{proof} $p^{(j)}_{01}\equiv p_j\alpha,\;p^{(j)}_{10}\equiv(1-p_j)\alpha$. (\Cref{fig:MarkovChain})
\[\pi^{(j)}_0p^{(j)}_{01}+\pi^{(j)}_1p^{(j)}_{11}=\pi^{(j)}_1\implies\pi^{(j)}_0p^{(j)}_{01} =\pi^{(j)}_1p^{(j)}_{10}\] \[\implies\pi^{(j)}_0\alpha p_j=\pi^{(j)}_1\alpha(1-p_j)\]
\[\implies p_j=\pi^{(j)}_1=\text{E}N^{(j)}_1/(N^{(j)}_1+N^{(j)}_0)=\text{E}N^{(j)}_1/s\]
\[\implies\text{E}\hat p_{j\text{eq}}=\text{E}N^{(j)}_1/s=p_{j}\implies \hat p_{j\text{eq}}\text{ is unbiased}\]
\end{proof}
\vspace{-0.3cm}

The above propositions imply that the larger the window size the better the estimation, and that in equilibrium, 
the temporal dependency (continuity rate) has no impact on estimating the onset probability of a Markov chain, 
and hence no impact on estimating the edge probability of a snapshot.

\begin{figure}[!thb]
\hspace{-00pt}
\begin{minipage}{240pt}
\includegraphics[width=240pt,height=90pt]{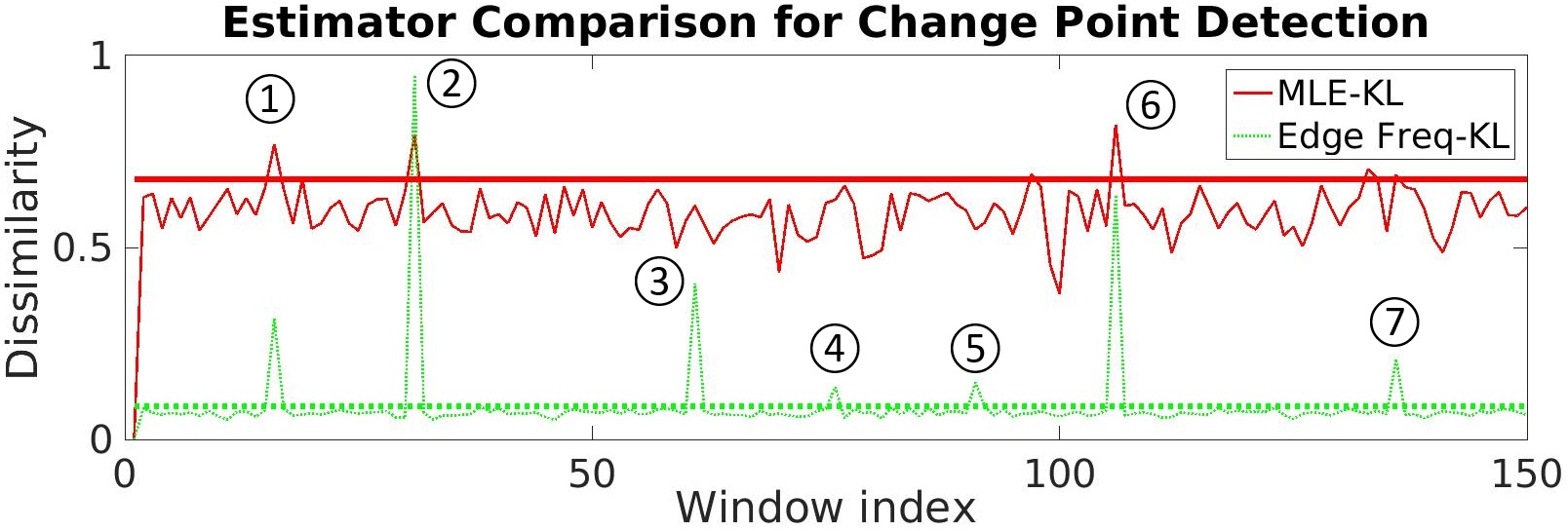} 
\captionsetup{width=240pt}
\caption{\footnotesize{Comparison between MLE (red, top) and Simplified Estimator (green, bottom) for change point detection on the same sequence as \Cref{fig:SBMResult} and \Cref{table:SBM_Exp}. Distance measure KL is used and other measures have consistent results; horizontal bars are corresponding thresholds. MLE has large fluctuation, and increasing window size reduces fluctuation; MLE misses true changes at \textcircled{3}, \textcircled{4} and \textcircled{5}, while edge frequency estimator has perfect recall and precision (see \Cref{sect:Experiment} for detail). (re-scaled and shifted for visualization)}}  
\label{fig:MLEFreqCompare}
\end{minipage}
\vspace{-0.1cm}
\end{figure}

Although MLE is close to the true value when the chain is long enough ($100$ or longer), 
we do not use so large a window size in practice ($20$ usually, no larger than $50$). 
Experiments (\Cref{fig:MLEFreqCompare}) show that the simplified estimator is much better than MLE for change point detection in practice.

\subsection{Distance measure} 
\label{sect:DistanceM}
Now, we need to compare the probability distributions of edges across consecutive windows. 
Kolmogorov-Smirnov (KS) statistic and Kullback-Leibler (KL) divergence are two common measures for comparing distribution. 
Their calculations require the enumeration of the whole state space and hence exponential to the number of variables for joint distributions. 
Although KS statistic is designed for univariate distribution, 
we can map the joint distribution, which has multivariate binary variables, 
to one dimension by decoding the binary vectors as an integer and use KS statistic. 
We bootstrap from empirical distributions of two consecutive windows respectively 
and use two sample KS test to quantify the difference of two distributions. 
We can use divide-and-conquer to alleviate the exponential complexity: 
partition the dyads into $g$ groups, compute KL/KS dissimilarity within each small group, 
and record the median among all the $g$ groups as the final dissimilarity.

Both of the above measures have good quality in terms of change point detection (\Cref{fig:SBMResult}), 
but KS statistic is extremely slow (\Cref{table:Time}), mostly due to the large sample bootstrap from each window. 
Euclidean distance, though lack of probability interpretation, has linear complexity and has reasonable quality in practice. 

\begin{table*}[!ht]
\vspace*{-0.2cm}
\small
\center
\begin{threeparttable}
\caption{\small{Model Change Explanation for an SBM-CL Experiment in \Cref{fig:SBMResult}}}
\label{table:SBM_Exp}
%\hspace*{-0.15in}
\begin{tabular}{p{25pt}p{55pt}p{390pt}}
\hline\hline
Order & Window Index & Type of Change\\
\txtcircled{1} & 15 & The weight sequence of 1/3 of the nodes is re-generated\\
\txtcircled{2} & 30 & The weight sequence of 2/3 of the nodes is re-generated\\
\txtcircled{3} & 60 & Half of the communities change their (inter- and intra-community) connection rate, overall density retained\\
\txtcircled{4} & 75 & All of the communities change their (inter- and intra-community) connection rate, overall density retained\\
\txtcircled{5} & 90 & Half of the communities change their (inter- and intra- community) connection rate, overall density changed\\
\txtcircled{6} & 105 & All of the communities change their (inter- and intra- community) connection rate, overall density changed\\
\txtcircled{7} & 135 & Community assignments of all the nodes are changed\\
\hline
\end{tabular}
\end{threeparttable}
\vspace*{-0.1cm}
\end{table*}

\subsection{Threshold Determination} 
\label{subsect:Threshold}
Suppose we have $w$ windows, then we compare $w-1$ pairs of distributions and get $w-1$ difference/distance scores. 
How do we choose a threshold to determine at which window the network changes? 
We use a permutation test \cite{pitman1937significance} based approach to determine the threshold. 
For a desired significance level $\alpha_s$, we bootstrap from the $w-1$ distance scores, 
and use the upper $100\alpha_s\%$ quantile as the threshold. 

\subsection{Complexity Analysis}
The algorithm is linear to the number of windows and constant to the network size for moderately large network.
Only a small fraction of dyads in the network is sampled and tracked. 
The sampling of the dyads is only performed once at the beginning, 
and hence irrelevant to the number of snapshots. 
For each snapshot, selecting a specific set of dyads has linear cost to the number of edges. 
Each window is only scanned once and therefore the time cost is linear to the number of windows. 
Moreover, since the number of windows is linear to the total number of snapshots 
even in the worst case (windows are overlapping, and window step is one), 
the algorithm is linear to the number of snapshots. 
Therefore, the time complexity is $O(\bar{M}T)$, 
where $\bar{M}$ is the averaged number of edges in each snapshot, 
and $T$ is the number of snapshots.

The memory cost is low, and can be viewed as constant: 
for each snapshot, only the information of the tracked dyads is stored;
information of dyads within the same window is aggregated; 
dyads information in the old window is overwritten once it is compared against the new window.
And the space complexity is $O(c)$, where $c$ is a prescribed sample size. 
Theoretically the sample size should be proportional to the network size for good estimation. 
Our experiments show that a fixed sample size (to track 250 out of $1.2$G dyads 
$\approx\binom{50\text{k}}{2}$) works well on a moderately large network.

\begin{figure}[h!]
\vspace{-0.2cm}
\hspace{-0.4cm}
\begin{minipage}{240pt}
\subfloat[Likelihood (ground truth)\label{fig:LikelihoodSBM}]{\includegraphics[width=240pt,height=140pt]{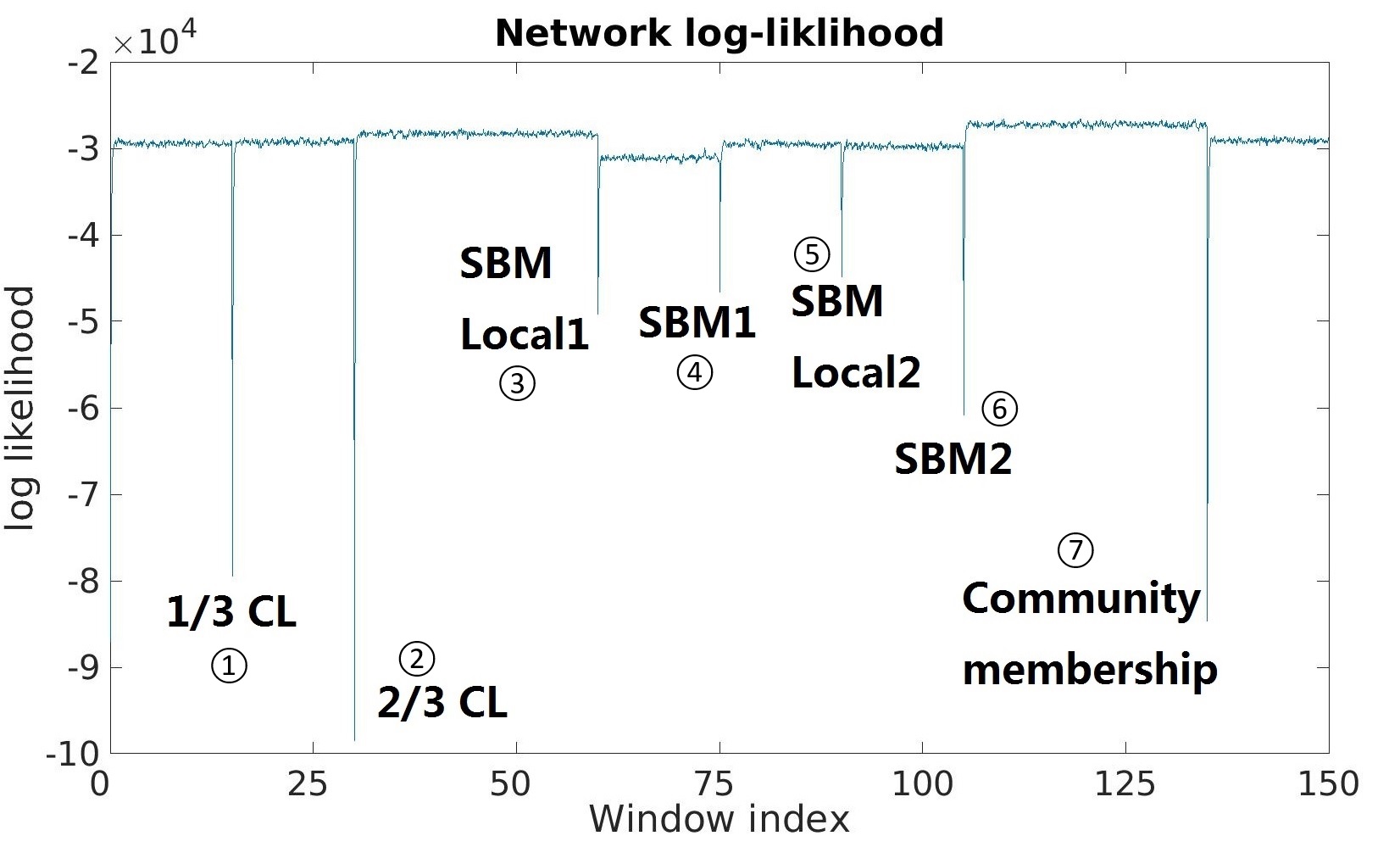}}\hfill
\vspace{-0.0cm}
\subfloat[Algorithms comparison. Curves are dissimilarity scores and horizontal bars are thresholds, and they two have corresponding colors and line shapes. (re-scaled and shifted for visualization)\label{fig:DeltaConSBM}]{\includegraphics[width=240pt,height=150pt]{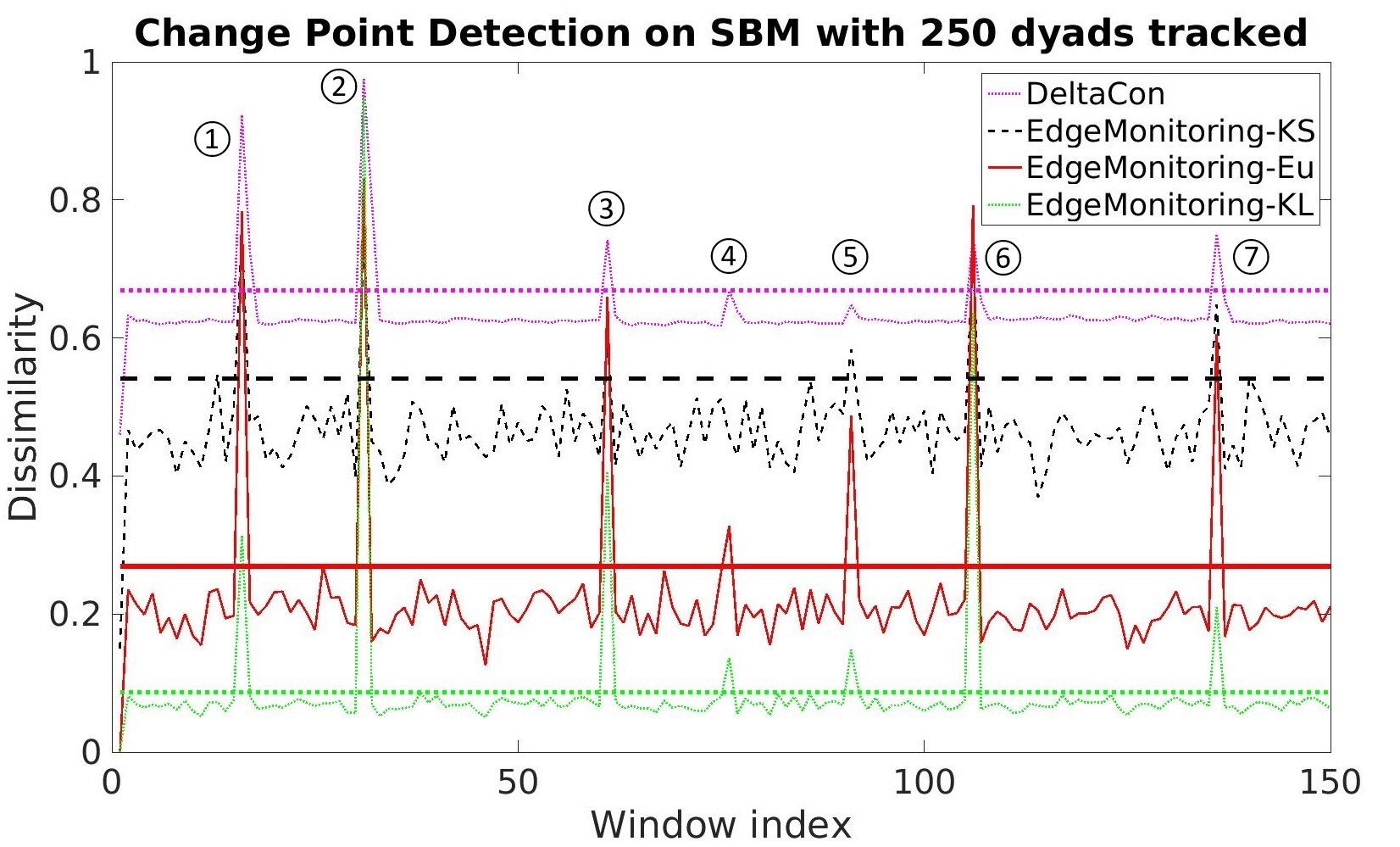}}
\end{minipage}
\captionsetup{width=240pt}
\hspace{40pt}
\caption{\small{SBM, ground truth changes explained in \Cref{table:SBM_Exp}. $1-\alpha$ = 0.51 and window size = 20. DeltaCon (Fig b-top) and EM-KL (Fig b-bottom) have the smallest variance, but DeltaCon has two false negatives at \textcircled{4} and \textcircled{5}.}}
\label{fig:SBMResult}
\end{figure}

\section{Experiments And Results}
% One figure has ground truth likelihood, the other has KS, KL, Eu and DeltaCon.
\label{sect:Experiment}
\newcommand{\ifNotGBTER}{\iffalse}

We did thorough evaluation of our edge probability estimation based 
change point detection algorithm (called EdgeMonitoring for simplicity) 
on synthetic and real world datasets. 
For the synthetic datasets, the generative process is known, 
and we can compute the ground truth in the form of likelihood, 
which is naturally a baseline choice. 
We also use the state-of-the-art DeltaCon~\cite{koutra2016deltacon} and LetoChange~\cite{peel2014detecting} as two baselines. 

\begin{figure}[!th]
\vspace{-0.2cm}
\hspace{-0.1cm}
\begin{minipage}{240pt}
\includegraphics[width=240pt,height=130pt]{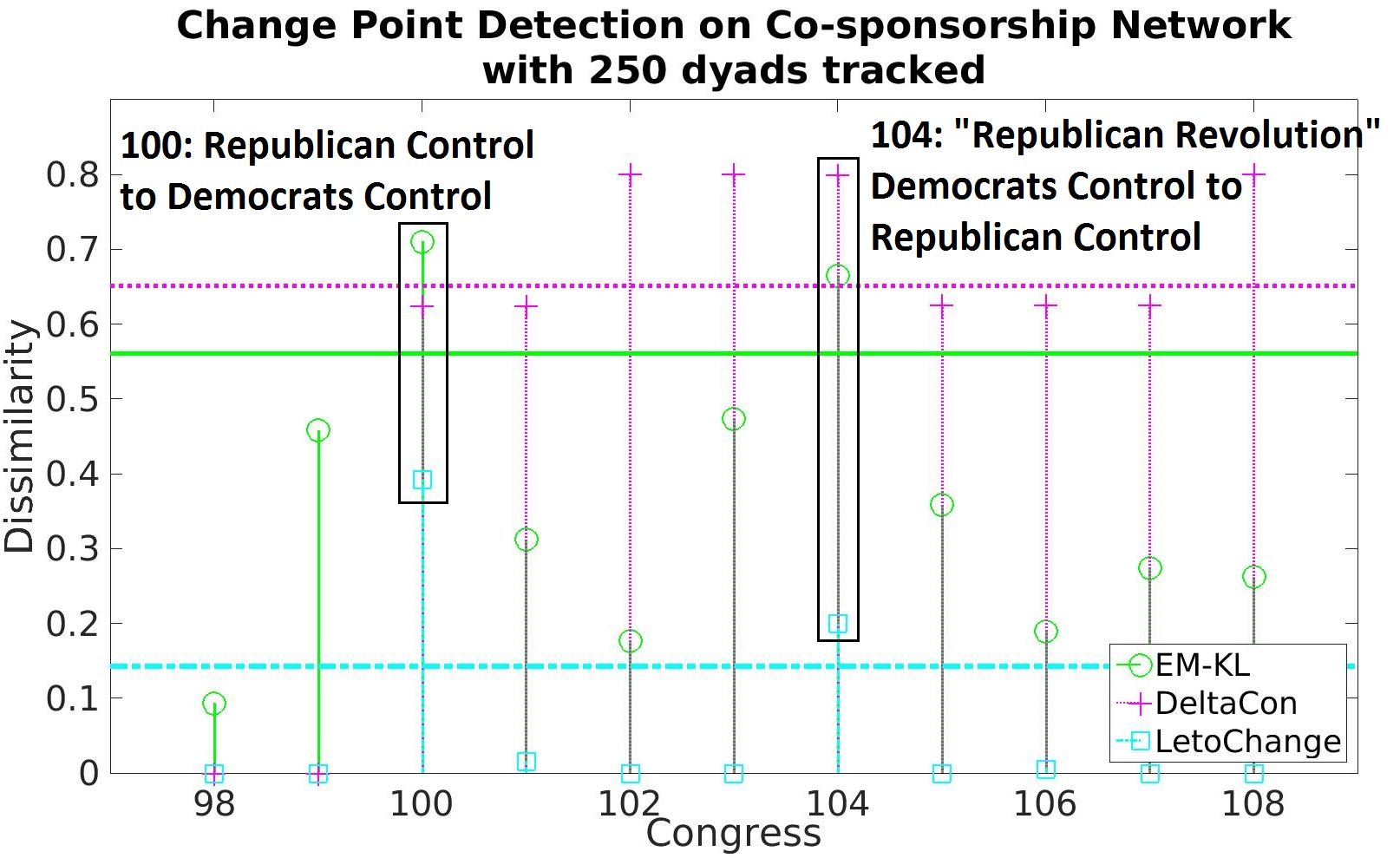}
\captionsetup{width=240pt}
\vspace{-0.7cm}
\caption{\small{Change point detection on US Senate co-sponsorship network. Change points at the 100th and the 104th Congresses (boxed) correspond to partisan domination shifts. Both EM-KL (green) and LetoChange (cyan) have perfect recall and precision, while DeltaCon (pink) has 3 false positives and 1 false negative.}}
\label{fig:CongressNetwork}
\end{minipage}
\vspace{-0.3cm}
\end{figure}

\subsection{Synthetic Data}
\textbf{Data generation}\footnote{\footnotesize{Generated using SNAP\cite{snap}}} We generate a sequence of networks from a fixed generation model.
The snapshots are not independent, each snapshot depends on the preceding
one through the continuity parameter $\alpha(\alpha_t\equiv\alpha)$. For each snapshot, each edge is selected
independently with probability $\alpha$, and if selected, the edge is again sampled from the
generative model (Table \ref{table:edgeP}). We introduce the change points
by changing the generative model in the middle of the sequence of snapshots. 
Note that this change may be simply a change of parameter values for a given model (Eg. ER$0.4$ to ER$0.6$),
or a change in the model type (Eg. SBM to ER), as well. 
Since our algorithm makes no assumptions about model specifics, 
we are able to detect both kinds of changes. 
We only inject parameter change in the synthetic experiment since the latter change is easily detectable. 
Sample changes are displayed in \Cref{table:SBM_Exp}. 
The likelihood of the snapshot sequence is also provided. 

We ran experiments with network sizes ranging from 1k to 50k, window size to
be from 10 to 100 and continuity rate $1-\alpha$ to be 0.51 and 0.9. We generated a total of 
5000 snapshots and sampled 250 edges uniformly at random to track. 
Both overlapping window ($s$$=$$2\eta$) and non-overlapping window have similar results, 
yet the latter is faster simply due to fewer windows. 
Hence we display non-overlapping window results only. 
For KL and KS, edges are grouped into 25 equal-sized groups. 
We use upper $5\%$ quantile as the threshold.% \looseness=-1

\textbf{Results} 
\Cref{fig:SBMResult} shows the qualitative comparison and \Cref{table:Time} reports the efficiency. 
\Cref{fig:LikelihoodSBM} shows the likelihood of the network drops 
dramatically after the generative model changes, and recovers to new equilibrium afterwards. 
Our EdgeMonitoring (EM-Eu, EM-KL) approach can successfully identify all change points with 5X speed up over DeltaCon. 
The changes are explained in \Cref{table:SBM_Exp}. 
We can see that EM-KL has the best performance: 
little fluctuation and perfect precision and recall. 
DeltaCon, though has smallest fluctuation, misses two change points. 
Both EM-KS and EM-Eu have large fluctuation. 
The quality of EM-KS heavily relies on the joint probability estimation, and we do see smaller fluctuation and higher recall for larger window size. 
EM-Eu in general has large fluctuation. 
EM-KL has the best overall performance, in terms of both quality and time efficiency. 
We believe grouping together with median selection contribute to its superiority.

\begin{table}[!ht]
%\vspace*{-0.2cm}
\footnotesize
%\center
\caption{\small{Time efficiency comparison (5k snapshots)}}
\label{table:Time}
\vspace{-0.1cm}
\hspace{-0.2cm}
\begin{threeparttable}
\begin{tabular}{p{34pt}p{26pt}p{26pt}p{23pt}p{28pt}p{36pt}p{26pt}}%{ccccccc}
\hline\hline
\hspace{-2pt}Model & Network Size & Window Size & EM Time\tnote{1} & EM-KS Time & DC Time (speedup) & LC Time\\
\hspace{-2pt}CL  & 1k & 20 & \textbf{18s} & 11h & \textcolor{white}{0}91s (5X) & DNF\\
\hspace{-2pt}SBM-CL  & 1k & 10 & \textbf{27s} & 22h & 125s (5X) & DNF\\
\hspace{-2pt}SBM-CL  & 1k & 50 & \textbf{9s} & 4.5h & \textcolor{white}{0}43s (5X) & DNF\\
\hspace{-2pt}SBM-CL  & 5k & 20 & \textbf{54s} & 11h & 309s (6X) & DNF\\
\hspace{-2pt}SBM-CL  & 10k & 20 & \textbf{232s} & 10h & \textcolor{white}{.}32m(8X) & DNF\\
\hspace{-2pt}SBM-CL  & 50k & 20 & \textbf{26m} & 10h & \textcolor{white}{i0}4h (9X) & DNF\\
\hspace{-2pt}BTER\tnote{2} & 1k & 20 & \textbf{3s} & 87m & \textcolor{white}{0}12s (4X) & 6h\\
\hspace{-2pt}\Cref{fig:SBMResult} & 1k & 20 & \textbf{21}s & 6.5h & 103s (5X) & DNF\\
\hspace{-2pt}\Cref{fig:CongressNetwork} & 100 & biennial & \textbf{4}s & 43m & \textcolor{white}{0}16s (4X) & 13h\\
\hspace{-2pt}\tiny{\cite{voeten2012data}} & $\approx$200 & annual & \textbf{10}s & 3h & \textcolor{white}{0}93s (9X) & DNF\\
\hspace{-2pt}Enron & 150 & weekly & \textbf{1}s & 7.5h & \textcolor{white}{00}1s (1X) & 60h\\
\hline
\end{tabular}
\begin{tablenotes}
\item[1]{EM for EdgeMonitoring (running time includes both KL and Euclidean), EM-KS for EdgeMonitoring with KS test, DC for DeltaCon, LC for LetoChange. EM and DC are implemented in MATLAB while LC in Python. All run on a commercial desktop with 48hrs as time limit. Each running time averaged over 5 runs.}
\item[2]{BTER dataset has 800 snapshots}
\end{tablenotes}
\end{threeparttable}
%\vspace{-0.5cm}
\end{table}

\subsection{Real World Data}
\textbf{Senate cosponsorship network}(\cite{fowler2006legislative}) 
We construct a co-sponsorship network from bills (co-)sponsored in US Senate during the 93rd-108th Congress. 
An edge is formed between two congresspersons if they cosponsored the same bill. 
Each bill corresponds to a snapshot, and forms a clique of co-sponsors. 
A window is set to include all bills in a single Congress (Biennially).

We randomly selected 250 dyads and tracked their fluctuations across the Congresses. 
We start from the 97th Congress since full amendments data is available only from 97th session onwards. 
\Cref{fig:CongressNetwork} compares EdgeMonitoring+KL, DeltaCon and LetoChange.
All methods were able to detect the most significant change point at the 104th Congress. 
Fowler \cite{fowler2006legislative} points out that there was a ``Republican Revolution'' in the 104th Congress which ``\textit{caused a dramatic change in the partisan and seniority compositions}.'' 
The author also points out the significance of the 100th (highest clustering coefficient, 
significant collaboration) and 104th Congress (lowest clustering coefficient, low point in collaboration) 
as inflection points in the Senate political process. 
Both our EdgeMonitoring approach and LetoChange classify these two Congresses as change points, but the latter takes much more time. 
DeltaCon picks up on one (104th) and not the other (100th). 
This provides evidence that our algorithm is able to capture the changes in 
network evolution effectively while being significantly faster than the state-of-the-art.

\section{Conclusion}
\label{sect:Conclusion}
In this paper, we develop a change point detection algorithm for dynamic networks that is efficient and accurate. 
Our approach relies on sampling and comparing the estimated joint edge (dyad) distribution. 
We first develop a maximum likelihood estimator, and analyze its drawbacks for small window sizes (the typical case). 
We then develop a consistent and unbiased estimator that overcomes the drawbacks of the MLE, resulting in significant quality improvement over the MLE. 
We conduct a thorough evaluation of our change point detection algorithm against two state-of-the-art DeltaCon and LetoChange on synthetic as well as the real world datasets. 
Our results indicate that our method is up to 9X faster than DeltaCon while achieving better quality.
\iffalse
In the future we plan to examine if we can employ a progressive refinement approach 
(by lowering the threshold) making the algorithm more sensitive, resulting high recall but potentially at the cost of false positives. 
One can then run a more expensive change point detection algorithm as the number of potential candidates become reasonably small after the first pass. 
Such an approach may yield more robust results while still being computationally efficient. 
\fi
%Moreover, we would like to try ensemble detector, that is, to combine the results from multiple change point detection algorithms. 
%This should also give us better quality. 
In the future we plan to extend our work to track higher order structures of 
the network such as 3-profiles~\cite{elenberg2015beyond} or 4-profiles and see how they evolve over time.

\section*{Acknowledgments}
This work is supported in part by NSF grant DMS-1418265, IIS-1550302 and IIS-1629548. 
Any opinions, findings, and conclusions or recommendations expressed in this material 
are those of the authors and do not necessarily reflect the views of the National Science Foundation.

%% The file named.bst is a bibliography style file for BibTeX 0.99c
\bibliographystyle{named}
\bibliography{myRef}

\begin{thebibliography}{}

\bibitem[\protect\citeauthoryear{Akoglu \bgroup \em et al.\egroup
  }{2014}]{AkogluTK14}
Leman Akoglu, Hanghang Tong, and Danai Koutra.
\newblock Graph-based anomaly detection and description: {A} survey.
\newblock {\em Data Mining and Knowledge Discovery (DAMI)}, 28(4), 2014.

\bibitem[\protect\citeauthoryear{Berlingerio \bgroup \em et al.\egroup
  }{2012}]{berlingerio2012netsimile}
Michele Berlingerio, Danai Koutra, Tina Eliassi-Rad, and Christos Faloutsos.
\newblock Netsimile: a scalable approach to size-independent network
  similarity.
\newblock {\em arXiv preprint arXiv:1209.2684}, 2012.

\bibitem[\protect\citeauthoryear{Bridges \bgroup \em et al.\egroup
  }{2015}]{bridges2015multi}
Robert~A Bridges, John~P Collins, Erik~M Ferragut, Jason~A Laska, and Blair~D
  Sullivan.
\newblock Multi-level anomaly detection on time-varying graph data.
\newblock In {\em Proceedings of the 2015 IEEE/ACM International Conference on
  Advances in Social Networks Analysis and Mining 2015}, pages 579--583. ACM,
  2015.

\bibitem[\protect\citeauthoryear{Caceres and
  Berger-Wolf}{2013}]{caceres2013temporal}
Rajmonda~Sulo Caceres and Tanya Berger-Wolf.
\newblock Temporal scale of dynamic networks.
\newblock In {\em Temporal Networks}, pages 65--94. Springer, 2013.

\bibitem[\protect\citeauthoryear{Elenberg \bgroup \em et al.\egroup
  }{2015}]{elenberg2015beyond}
Ethan~R Elenberg, Karthikeyan Shanmugam, Michael Borokhovich, and Alexandros~G
  Dimakis.
\newblock Beyond triangles: A distributed framework for estimating 3-profiles
  of large graphs.
\newblock In {\em Proceedings of the 21th ACM SIGKDD International Conference
  on Knowledge Discovery and Data Mining}, pages 229--238. ACM, 2015.

\bibitem[\protect\citeauthoryear{Erd\H{o}s and
  R{\'e}nyi}{1960}]{erd6s1960evolution}
Paul Erd\H{o}s and A~R{\'e}nyi.
\newblock On the evolution of random graphs.
\newblock {\em Publ. Math. Inst. Hungar. Acad. Sci}, 5:17--61, 1960.

\bibitem[\protect\citeauthoryear{Fowler}{2006}]{fowler2006legislative}
James~H Fowler.
\newblock Legislative cosponsorship networks in the {US} {H}ouse and {S}enate.
\newblock {\em Social Networks}, 28(4):454--465, 2006.

\bibitem[\protect\citeauthoryear{Givens and
  Hoeting}{2012}]{givens2012computational}
Geof~H Givens and Jennifer~A Hoeting.
\newblock {\em Computational statistics}, volume 710.
\newblock John Wiley \& Sons, 2012.

\bibitem[\protect\citeauthoryear{Hunter \bgroup \em et al.\egroup
  }{2012}]{hunter2012computational}
David~R Hunter, Pavel~N Krivitsky, and Michael Schweinberger.
\newblock Computational statistical methods for social network models.
\newblock {\em Journal of Computational and Graphical Statistics},
  21(4):856--882, 2012.

\bibitem[\protect\citeauthoryear{Karrer and
  Newman}{2011}]{karrer2011stochastic}
Brian Karrer and Mark~EJ Newman.
\newblock Stochastic blockmodels and community structure in networks.
\newblock {\em Physical Review E}, 83(1):016107, 2011.

\bibitem[\protect\citeauthoryear{Klimt and Yang}{2004}]{klimt2004enron}
Bryan Klimt and Yiming Yang.
\newblock The enron corpus: {A} new dataset for email classification research.
\newblock In {\em Machine learning: ECML 2004}, pages 217--226. Springer, 2004.

\bibitem[\protect\citeauthoryear{Koutra \bgroup \em et al.\egroup
  }{2016}]{koutra2016deltacon}
Danai Koutra, Neil Shah, Joshua~T Vogelstein, Brian Gallagher, and Christos
  Faloutsos.
\newblock Deltacon: {P}rincipled {M}assive-{G}raph {S}imilarity {F}unction with
  {A}ttribution.
\newblock {\em ACM Transactions on Knowledge Discovery from Data (TKDD)},
  10(3):28, 2016.

\bibitem[\protect\citeauthoryear{La~Fond \bgroup \em et al.\egroup
  }{2014}]{la2014anomaly}
Timothy La~Fond, Jennifer Neville, and Brian Gallagher.
\newblock Anomaly detection in networks with changing trends, 2014.

\bibitem[\protect\citeauthoryear{{L}eskovec and {R}ok {S}osi\v{c}}{2014}]{snap}
Jure {L}eskovec and {R}ok {S}osi\v{c}.
\newblock {SNAP}: {A} general purpose network analysis and graph mining library
  in {C++}.
\newblock \url{http://snap.stanford.edu/snap}, Jun 2014.

\bibitem[\protect\citeauthoryear{Li \bgroup \em et al.\egroup
  }{2016}]{li2016detecting}
Shuang Li, Yao Xie, Mehrdad Farajtabar, and Le~Song.
\newblock Detecting weak changes in dynamic events over networks.
\newblock {\em arXiv preprint arXiv:1603.08981}, 2016.

\bibitem[\protect\citeauthoryear{Loglisci \bgroup \em et al.\egroup
  }{2015}]{loglisci2015relational}
Corrado Loglisci, Michelangelo Ceci, and Donato Malerba.
\newblock Relational mining for discovering changes in evolving networks.
\newblock {\em Neurocomputing}, 150:265--288, 2015.

\bibitem[\protect\citeauthoryear{mat}{}]{mathematica}
Wolfram mathematica.
\newblock \url{https://www.wolfram.com/mathematica/}.
\newblock Accessed: 2017-06-03.

\bibitem[\protect\citeauthoryear{Moreno and Neville}{2013}]{moreno2013network}
Sebastian Moreno and Jennifer Neville.
\newblock Network hypothesis testing using mixed kronecker product graph
  models.
\newblock In {\em Data Mining (ICDM), 2013 IEEE 13th International Conference
  on}, pages 1163--1168. IEEE, 2013.

\bibitem[\protect\citeauthoryear{Peel and Clauset}{2015}]{peel2014detecting}
Leto Peel and Aaron Clauset.
\newblock Detecting change points in the large-scale structure of evolving
  networks.
\newblock In {\em Twenty-Ninth AAAI Conference on Artificial Intelligence},
  2015.

\bibitem[\protect\citeauthoryear{Peixoto and
  Rosvall}{2015}]{peixoto2015modeling}
Tiago~P Peixoto and Martin Rosvall.
\newblock Modeling sequences and temporal networks with dynamic community
  structures.
\newblock {\em arXiv preprint arXiv:1509.04740}, 2015.

\bibitem[\protect\citeauthoryear{Pfeiffer~III \bgroup \em et al.\egroup
  }{2012}]{pfeiffer2012fast}
Joseph~J Pfeiffer~III, Timothy La~Fond, Sebastian Moreno, and Jennifer Neville.
\newblock Fast generation of large scale social networks with clustering.
\newblock {\em arXiv preprint arXiv:1202.4805}, 2012.

\bibitem[\protect\citeauthoryear{Pitman}{1937}]{pitman1937significance}
Edwin~JG Pitman.
\newblock Significance tests which may be applied to samples from any
  populations.
\newblock {\em Supplement to the Journal of the Royal Statistical Society},
  4(1):119--130, 1937.

\bibitem[\protect\citeauthoryear{Ranshous \bgroup \em et al.\egroup
  }{2015}]{ranshous2015anomaly}
Stephen Ranshous, Shitian Shen, Danai Koutra, Steve Harenberg, Christos
  Faloutsos, and Nagiza~F Samatova.
\newblock Anomaly detection in dynamic networks: a survey.
\newblock {\em Wiley Interdisciplinary Reviews: Computational Statistics},
  7(3):223--247, 2015.

\bibitem[\protect\citeauthoryear{Seshadhri \bgroup \em et al.\egroup
  }{2012}]{seshadhri2012community}
C~Seshadhri, Tamara~G Kolda, and Ali Pinar.
\newblock Community structure and scale-free collections of
  erd{\H{o}}s-r{\'e}nyi graphs.
\newblock {\em Physical Review E}, 85(5):056109, 2012.

\bibitem[\protect\citeauthoryear{Shih and
  Parthasarathy}{2012}]{shih2012identifying}
Yu-Keng Shih and Srinivasan Parthasarathy.
\newblock Identifying functional modules in interaction networks through
  overlapping markov clustering.
\newblock {\em Bioinformatics}, 28(18):i473--i479, 2012.

\bibitem[\protect\citeauthoryear{Voeten}{2012}]{voeten2012data}
Erik Voeten.
\newblock Data and analyses of voting in the {U}{N} general assembly.
\newblock {\em Available at SSRN 2111149}, 2012.

\bibitem[\protect\citeauthoryear{Zhang \bgroup \em et al.\egroup
  }{2016}]{zhang2016random}
Xiao Zhang, Cristopher Moore, and MEJ Newman.
\newblock Random graph models for dynamic networks.
\newblock {\em arXiv preprint arXiv:1607.07570}, 2016.

\end{thebibliography}

\end{document}